# Evolution of Landau Levels into Edge States at an Atomically Sharp Edge in Graphene.

Guohong Li[1], Adina Luican[1], Dmitry Abanin[2], Leonid Levitov[3], Eva Y. Andrei[1*]

[1]Department of Physics and Astronomy, Rutgers University, Piscataway NJ 08855, USA
[2] Department of Physics, Harvard University, Cambridge, MA 02138, USA
[3] Department of Physics, Massachusetts Institute of Technology, Cambridge, MA 02139, USA

**The quantum-Hall-effect (QHE) occurs in topologically-ordered states of two-dimensional (2d) electron-systems in which an insulating bulk-state coexists with protected 1d conducting edge-states. Owing to a unique topologically imposed edge-bulk correspondence these edge-states are endowed with universal properties such as fractionally-charged quasiparticles and interference-patterns, which make them indispensable components for QH-based quantum-computation and other applications. The precise edge-bulk correspondence, conjectured theoretically in the limit of sharp edges, is difficult to realize in conventional semiconductor-based electron systems where soft boundaries lead to edge-state reconstruction. Using scanning-tunneling microscopy and spectroscopy to follow the spatial evolution of bulk Landau-levels towards a zigzag edge of graphene supported above a graphite substrate we demonstrate that in this system it is possible to realize atomically sharp edges with no edge-state reconstruction. Our results single out graphene as a system where the edge-state structure can be controlled and the universal properties directly probed.**

Two-dimensional electron systems (2DES) in the Quantum-Hall (QH) regime host gapless one-dimensional chiral edge states near sample boundaries which are responsible for the quantization of the Hall conductivity(*1-3*). The chiral carriers in the edge states are either right or left-moving and reside on opposite sides of the sample. For well separated edges backscattering is suppressed leading to robust one-dimensional ballistic channels which are an ideal laboratory for the study of quantum transport in one dimension. In the regime of the fractional QHE these edge states are expected to form a new kind of strongly interacting non-Fermi-liquid, a chiral Luttinger liquid (CLL) (*4, 5*). The CLL reflects the topological structure of the underlying correlated electron state(*6-9*) and presents a palette of unusual properties including distinctive tunneling characteristics with power law singularities, shot-noise reflecting the fractional charge of the excitations, anionic or non-abelian statistics and interference patterns that could serve as building blocks, interconnects and probes for QH qubits. Observing and exploiting these properties requires precise control of the edges, but in semiconductor- based 2DES where edge states were studied thus far, achieving the necessary control proved challenging (*10, 11*). In these systems the lithographically defined edges have soft confinement potentials (caused by the gates and dopant layer being far away from the 2DES) and large boundary widths which can alter the edge states in unpredictable ways. In particular, when the boundary width exceeds the magnetic length, a series of compressible and incompressible strips forms (Fig. 1a) 'contaminating' the edge-bulk correspondence and the universal behavior (*12-15*). These features appear to be a key obstacle for many QH-based quantum computing applications.

Graphene, a one-atom thick crystal of C atoms arranged in a honeycomb lattice (*16*), provides unprecedented opportunities to revisit the physics of QH edge states. It is strictly 2d, exhibits a robust fractional QHE(*17, 18*), and the flexibility in the position of the gates and screening plane allows a systematic study of edge states under controlled screening conditions (*19, 20*). For gate-induced carriers the boundary width is determined by the distance to the gate, similar to other 2DES, but this can approach atomic length-scales as in the case of graphene on graphite (*21-23*) or SiC (*24*) . To date, despite extensive theoretical work (*25-30*) and several spectroscopic studies of LLs in the bulk (*21-24*)*,* quantum Hall edge states in graphene remain largely unexplored.

Here we employ scanning tunneling microscopy (STM) and spectroscopy (STS) to study the edge states in graphene on a graphite substrate. Previous work showed that graphite is minimally invasive and can provide access to the intrinsic electronic properties of graphene*(12-15)*. The fact that graphite is conducting, together with the ability to find graphene samples on it that are completely decoupled from the substrate (*22*), make it a perfect platform for STM/STS studies of edge states. Using STS to follow the spatial evolution of the LL peak sequence from the bulk towards an edge, we show that in the bulk the sequence of LLs is characteristic of massless Dirac fermions and that the special LL at the Dirac point (DP), with index *N* = 0, persists all the way to a zigzag edge. We find that the position of the LL peaks and the carrier density remain practically constant upon approaching the edge to within half a magnetic length. The bending of the LLs upon approaching the edge gives rise to redistribution of spectral weight from the peaks towards higher energies, which is clearly seen in the data and is in good agreement with the theory of a sharp edge (*27*). No evidence of compressible/incompressible strips or any other form of edge reconstruction is found, in sharp contrast to 2DES in semiconductor-based structures where the boundary width spans many magnetic lengths and edge reconstruction is inevitable (*31-35*). Finally, on the very edge we observe a dramatic change in topography and spectroscopy suggesting a new phenomenon associated with the edge termination.

The low energy band structure of graphene consists of electron-hole symmetric Dirac cones which touch at the DPs located at the K and K' corners (valleys) of the Brillouin zone. In a magnetic field *B* normal to the graphene plane the spectrum consists of discrete LLs:

$$E_n = E_D \pm \varepsilon_0 \sqrt{|N|} \qquad (1)$$

where *N* is the level index with *N* > 0 corresponding to electrons and *N* < 0 to holes, and $\varepsilon_0 = \sqrt{2e\hbar v_F^2 B}$ is a characteristic energy scale. Here $\hbar$ is the reduced Planck constant, $v_F \sim 10^6$ m/s the Fermi velocity, +/- refer to the electron/hole branches, and $E_D$ is the energy of the DP measured relative to the Fermi energy. The LL spectrum in graphene is qualitatively different from that of the 2DES in semiconductors: it is electron-hole-symmetric, displays square-root dependence on field and level index and it contains an *N* = 0 level which reflects the chiral nature of quasiparticles. The wave functions of the $N = 0$ LL in valleys K and K' reside on opposite sublattices, A and B, of the honeycomb lattice.

In the bulk the LLs are highly degenerate leading to pronounced peaks in the density of states (DOS). In the Landau gauge, natural to the strip geometry, the degeneracy reflects multiple choices for the position of the guiding-center and the wave-function forms extended strips parallel to the edge which straddle the guiding-centers over a width of the order of the magnetic length $l_B = \sqrt{\dfrac{\hbar}{eB}}$. For guiding-centers far from the edges the states are identical to those in an infinite system. Near an edge the wave function is modified owing to the boundary conditions at the edge, which results in LLs bending away from the DP. For both armchair and zigzag edges this lifts the valley degeneracy and leads to the formation of dispersive edge states that can carry transport currents. In addition, unique for the zigzag edge, there is a non-dispersive $N = 0$ state confined to one of the valleys.

Graphene flakes supported on a graphite substrate were studied using low-temperature high-magnetic-field scanning tunneling microscopy and spectroscopy (*22, 36*). Focusing on a region close to a zigzag edge (Fig.2a,b) the evolution of STS spectra was followed from bulk to the sample edge. Far from the edge the local density of states (LDOS) exhibits a series of pronounced peaks, with square root dependence on field and level index (Fig 2c,e), corresponding to a LL sequence characteristic of massless Dirac-fermions (Eq. 1). The splitting of the N=0 peak in the sequence can be attributed to a substrate induced staggered potential on the A and B sublattices which breaks the inversion symmetry and shifts the energies of the N=0 LL in K and K' valleys symmetrically about the DP. Far from the edge, for distances exceeding ~2.5 $l_B$, the LL sequence is insensitive to the proximity of the edge. Within closer distances to the edge we observe a redistribution of the spectral weight from the peaks to the valleys, but at the same time the positions of the peaks remain unchanged (Fig. 3). The double peak corresponding to the split $N = 0$ LL stands out in its robustness and persists all the way to the edge as expected for a zigzag termination (*25*). The fixed positions of the LL peaks all the way to the edge are the signature of an atomically sharp edge. In contrast, for a soft edge where the boundary-width is larger than the magnetic length as is the case of 2DES in semiconductors, bending of LLs would cause a shift in the peak energies upon approaching the edge. In order to quantify the spectral weight redistribution we plot in Fig. 3d the weight loss from peaks (on the electron side) as a function of distance from the edge and compare it to the weight gain in the valleys. The weight redistribution was obtained by subtracting the spectrum at $2.5 l_B$ from the spectrum at each position and calculating the area under the negative (positive) portions of the resultant curve to obtain the weight loss from peaks (gain from valleys). We find that the weight loss from the peaks is almost completely recovered by the gain in the valleys and that spectral weight shifts from low energy to higher energies in agreement with theory.

The energy of the DP ($E_D$), which is identified with the center of the two N=0 peaks with respect to the Fermi energy (defined as zero), measures the local carrier density $n = (E_D / \pi^{1/2} \hbar v_F)^{1/2}$ and sign. Far from the edge the sample is hole doped, $E_D > 0$, with a carrier density of n ~ $3 \times 10^{10}$ cm$^{-2}$. From the position dependence of the LL spectrum and $E_D$ upon approaching towards the zigzag edge we obtain in Fig. 3c the evolution of the local carrier density with distance from the edge. We note that the density remains practically unchanged upon approaching the edge to within ~1 $l_B$, again consistent with the absence of edge reconstruction expected for an atomically sharp edge.

For a quantitative comparison between theory and experiment we simulate the spatial evolution of the LDOS close to a zigzag edge including the level broadening due to the finite qausiparticle lifetime. Following Ref. (*27*), we use the low-energy continuum Dirac model to obtain LL energies, wave functions and the LDOS by solving numerically two Dirac equations in magnetic field (one per each valley), supplemented by the boundary condition appropriate for the zigzag edge. In addition, here we introduce splitting between N=0 Landau sub-levels by imposing different potentials $\pm\Delta$ on the two graphene sublattices. Furthermore, in order to account for the asymmetry of the split N=0 LL observed in the experiment, we assume that the tunneling matrix element into the two sublattices is different. This could arise from the asymmetric coupling of two sublattices to the graphite substrate. We found that taking $p_A = 2p_B$ (here $p_{A(B)}$ is the squared matrix element for tunneling into A(B) sublattice) gives the best agreement with experiment. Comparing to the measured LDOS in Fig. 3, we find that this simple model captures well the main experimental features, including the evolution of the LL peak heights with distance from the edge and the spectral weight redistribution (Fig.3b,d,f). Consistent with the experimental data the deviations from the bulk DOS appear only within ~2.5$l_B$ of the edge as a redistribution of intensity without shifting the positions of the LL peaks. Another notable feature, also consistent with experiment, is the persistence of the strong double-peak at the Dirac point all the way to the edge, even while the others are smeared out. Tellingly, because the state at the Dirac point persist in only one valley, the amplitude of one of the peaks in the N=0 doublet decreases for distances between 2.5 $l_B$ and 1.0 $l_B$ away from the edge.

The agreement with theory breaks down completely right on the edge, where the spectrum consists of three broad peaks unrelated to the bulk LLs ( Fig. 4a). At the same time atomic-resolution STM topography indicates a transition from honeycomb to triangular structure within ~0.5$l_B$ of the edge. The edge itself appears fuzzy but we note an unusual stripe pattern within the first few rows of atoms consistent with smearing of the triangular structure seen further away (Fig.4c). These new spectroscopic and topographic features at the zigzag edge of graphene cannot be understood within the existing theory. Density-functional calculations (*37*) of the charge distribution near a zigzag edge in zero field, exhibit a transition from honeycomb to triangular structure with a stripe-like termination on the edge similar to our observations. However, a closer look at Fig.4c reveals that the A sub-lattice is brighter while theory predicts that the B-sublattice should be brighter. Moreover the peculiar edge spectrum is also not accounted for by the theory. Since the samples were prepared under ambient conditions we must consider the possibility of adsorbed species at the edge, such as hydrogen or other carbohydrates. In the bulk, these adsorbates introduce a low energy midgap state within ~ 30meV of the DP (*38*) but their effect on the edge spectrum is not known. More work is needed to elucidate the role of edge termination, whether by reconstruction or accretion of adsorbates, on the spectrum. This alters the electronic and chemical properties of nano-graphene(*39*) and may have important implications for its potential use in application ranging from chemical sensors to electronics and metrology.

In conclusion, we have shown that when the screening plane is very close to the 2d electron layer as is the case for graphene on graphite, the QH edge states display the characteristics of confinement by an atomically sharp edge. The absence of edge reconstruction demonstrated here indicates that graphene is a suitable system for realizing one dimensional CLL states and for probing its universal properties as a projection of the underlying quantum Hall state. The findings reported here together with the techniques available to control the local density and the screening geometry in graphene, guarantee that edge softness and its undesirable reconstruction can be put under control in future experiments (this can be achieved, e.g., by using a combination of gating across a tunable gap in suspended samples and side gates (*40*). This new type of unreconstructed edge states will provide a test-bed for the theoretical ideas *(4, 5, 13, 19)* and will open new avenues for exploiting the physics of the 1d QH channels.

**Methods.** STM tips were mechanically cut from Pt-Ir wire. The tunneling conductance dI/dV was measured using lock-in detection at 340 Hz. A magnetic field, 4T, was applied perpendicular to sample surface. Typical tunneling junctions were set with 300mV sample bias voltage and 20 pA tunneling current. The samples were obtained from highly-oriented pyrolitic-graphite (HOPG) cleaved in air. In addition to removing surface contamination this methods often leaves graphene flakes on the graphite surface which are decoupled or weakly coupled to the substrate. The graphene flakes are characterized with topography followed by finite field spectroscopy in search of a well-defined and pronounced sequence of Landau levels indicating decoupling from the substrate (*22, 36*). We found that the strength of the peaks is a good first indication of the degree of coupling: the weaker the coupling of a flake to the substrate the stronger the peaks.

**Acknowledgements**

The authors acknowledge support from DOE DE-FG02-99ER45742 (EYA), NSF DMR-0906711 (GL), Lucent foundation (AL) and ONR N00014-09-1-0724 (LL).

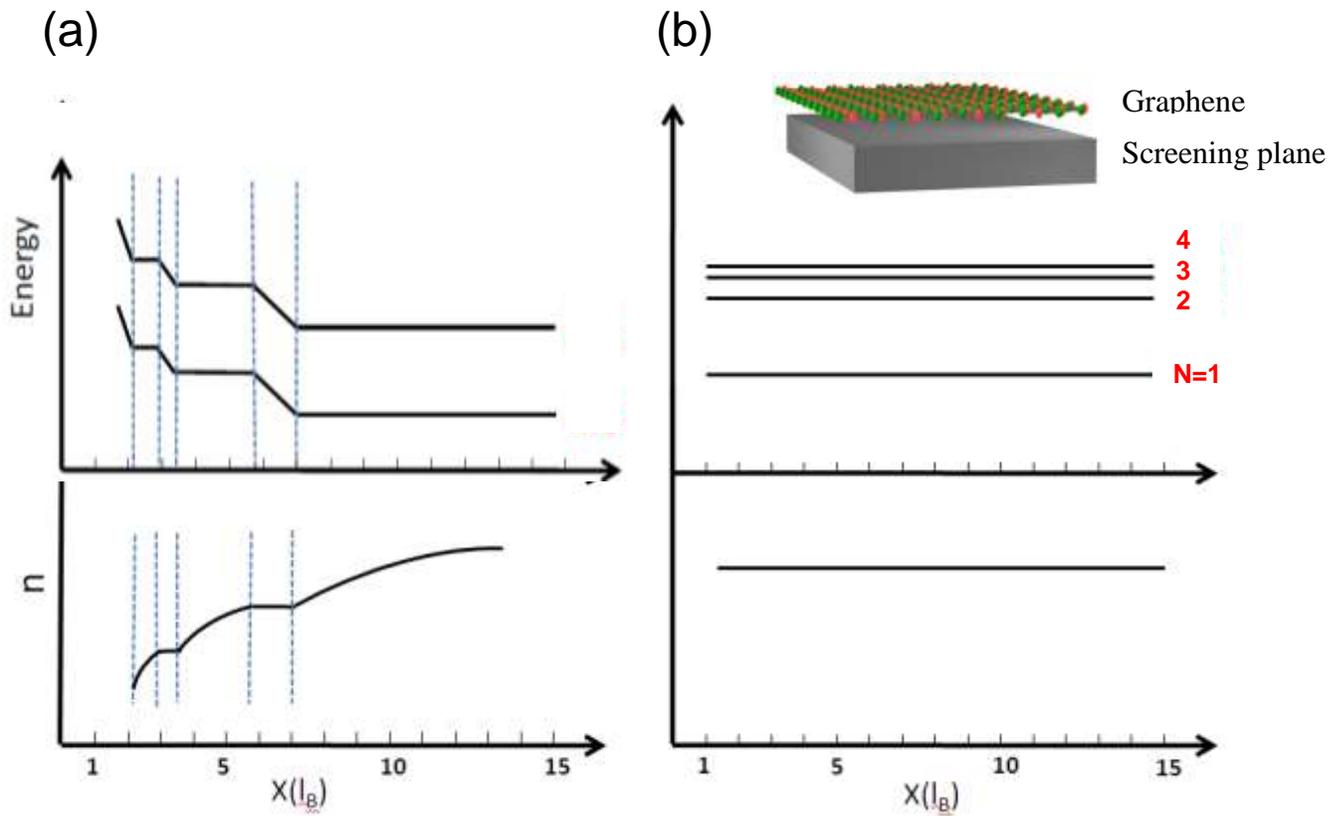

**Fig. 1. Effect of geometry and screening plane distance on edge states in 2DES.**

(a) Edge reconstruction is semiconductor based 2DES. The distances from gates and screening-plane are much larger than the magnetic-length. Top: Spatial variation of Landau-level energy as a function of distance from the edge shows the effect of edge reconstruction. Dashed lines mark the boundary between compressible and incompressible strips. Bottom: Spatial variation of the reconstructed carrier density close to the edge. (b) Same as *a* for the case of graphene on graphite where the screening-plane distance is much smaller than the magnetic-length. No edge reconstruction occurs.

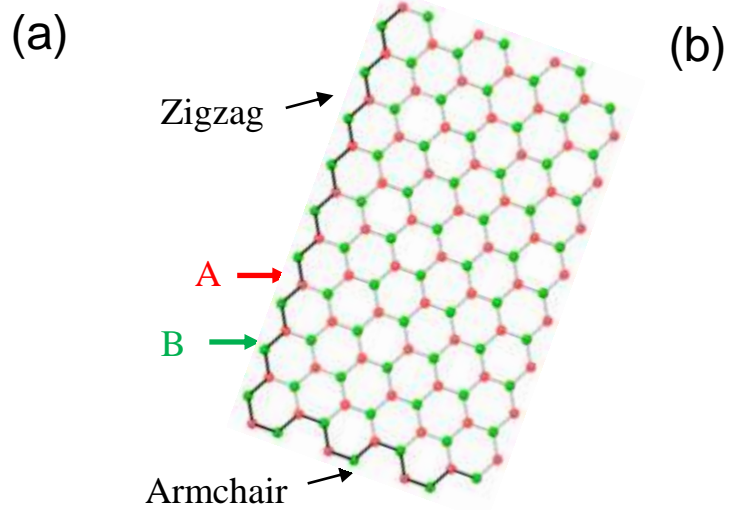
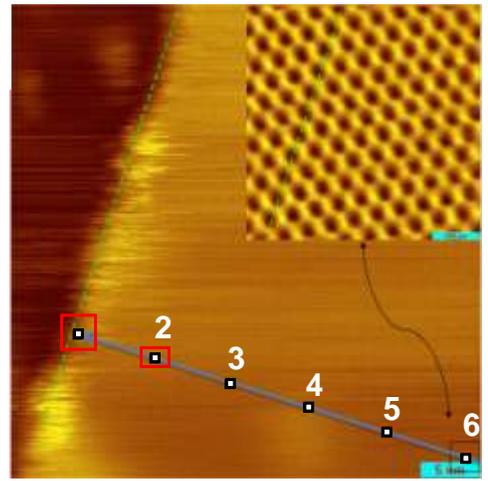
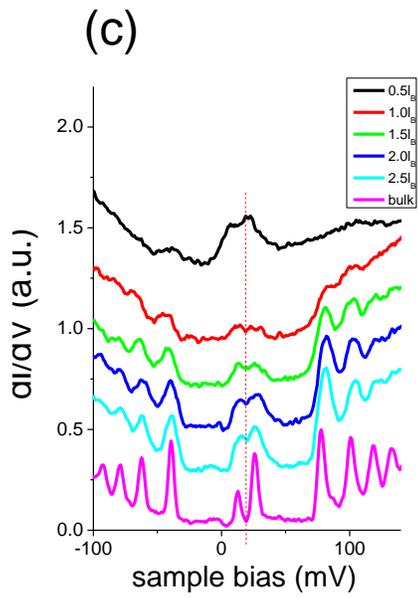
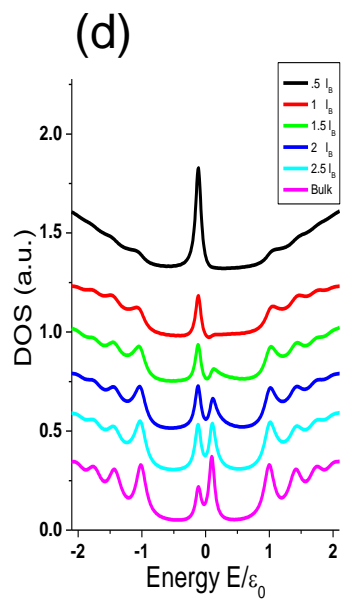
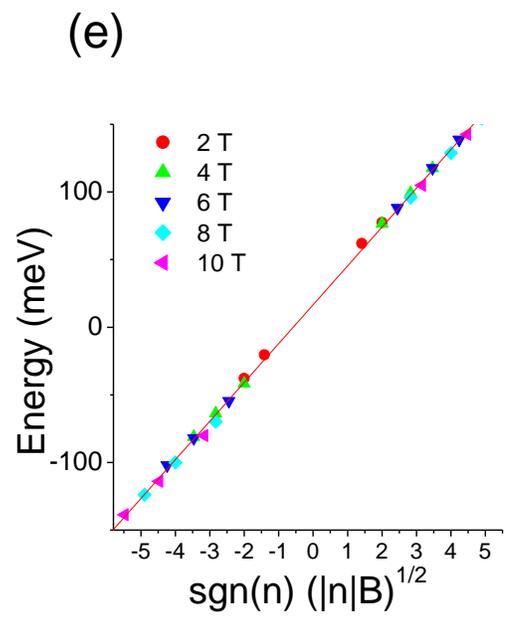

**Fig. 2. STM/STS on graphene near a zigzag edge.** (a) Graphene edges. The two sublattices in the honeycomb structure are denoted A and B. The zigzag edge termination contains either A or B-type atoms while the armchair contains both types. (b) STM of a graphene flake on graphite measured in a field of 4T at 4.4 K. Inset: the edge type is determined from atomic resolution STM at a distance of ~ 32nm = 2.5 $l_B$ from the edge (position 6 ) which shows a clear honeycomb structure. The dashed line marks a zigzag direction, and is parallel to the edge, also marked with a dashed line in the main panel. Spectra taken at intervals of 0.5 $l_B$ (marked 1-6) along the blue line are shown in panel *c*. Scale bars: 5nm (main panel), 500pm (inset). (c) STS at in the bulk and at the positions marked in panel *b*. The dashed line indicates the bulk Dirac point energy. (d) Simulated local density of states for the case in panel c including broadening due to electron-electron interactions obtained in reference (*22*). (e) Energy of LL sequence in the bulk versus the reduced parameter $(|N|B)^{1/2}$ indicates the massless Dirac fermion nature of the spectrum (equation 1).

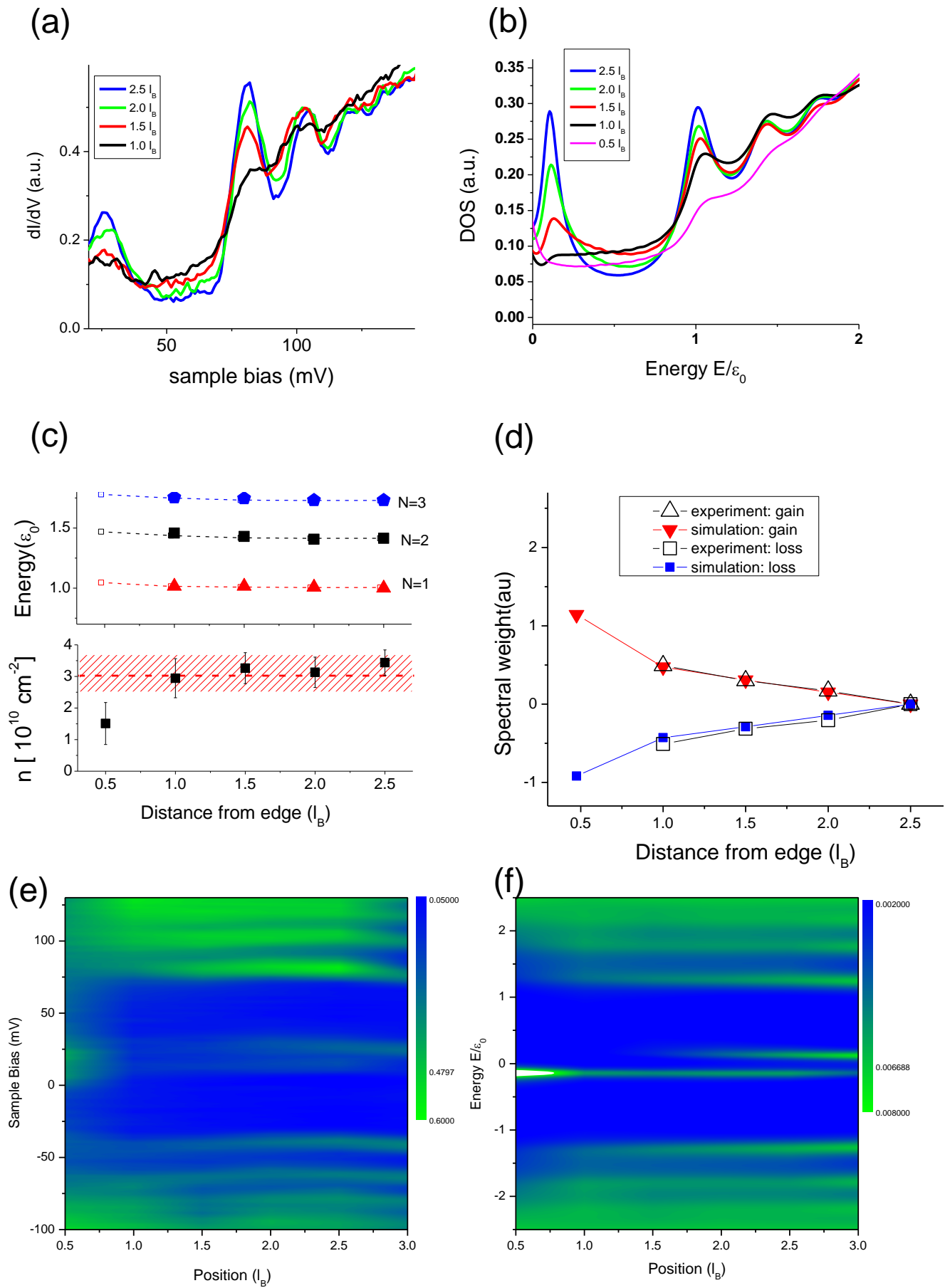

**Fig. 3. Evolution of LL with distance from the edge**. (a) Measured LL spectra showing the evolution of the peak heights on the electron side at B=4Tesla. (b) Simulated LDOS for the case in panel *a* including broadening due to electron-electron interactions obtained in reference 16. The LDOS is averaged over the two sub-lattices. Although the peaks corresponding to bulk LL continue to dominate the LDOS down to 0.5 $l_B$, their amplitude decreases upon approaching the edge. At the same time mid-gap states due to bending of the LLs contribute to the LDOS between the peaks. For distances exceeding $2.5 l_B$ the LDOS recovers the bulk LLs. (c) Top panel: comparison of measured LL peak position with distance from edge for N=1,2,3 (symbols) with calculated values (dashed lines) for a sharp edge show good agreement. Bottom panel: evolution of carrier density with distance from the edge shows no charge variation up to a distance of 0.5 $l_B$. (d) Spectral weight redistribution near the edge. Measured spectral weight loss from peaks on the electron side (solid squares) compared with the calculated values (open squares). Measured spectral weight gain in valleys (solid stars) coincides with the calculated values (open stars). (e) Measured LL maps at B=4 Tesla showing the evolution of the spectra with distance from edge. (d) Simulated LL maps for the same parameters as in *c*.

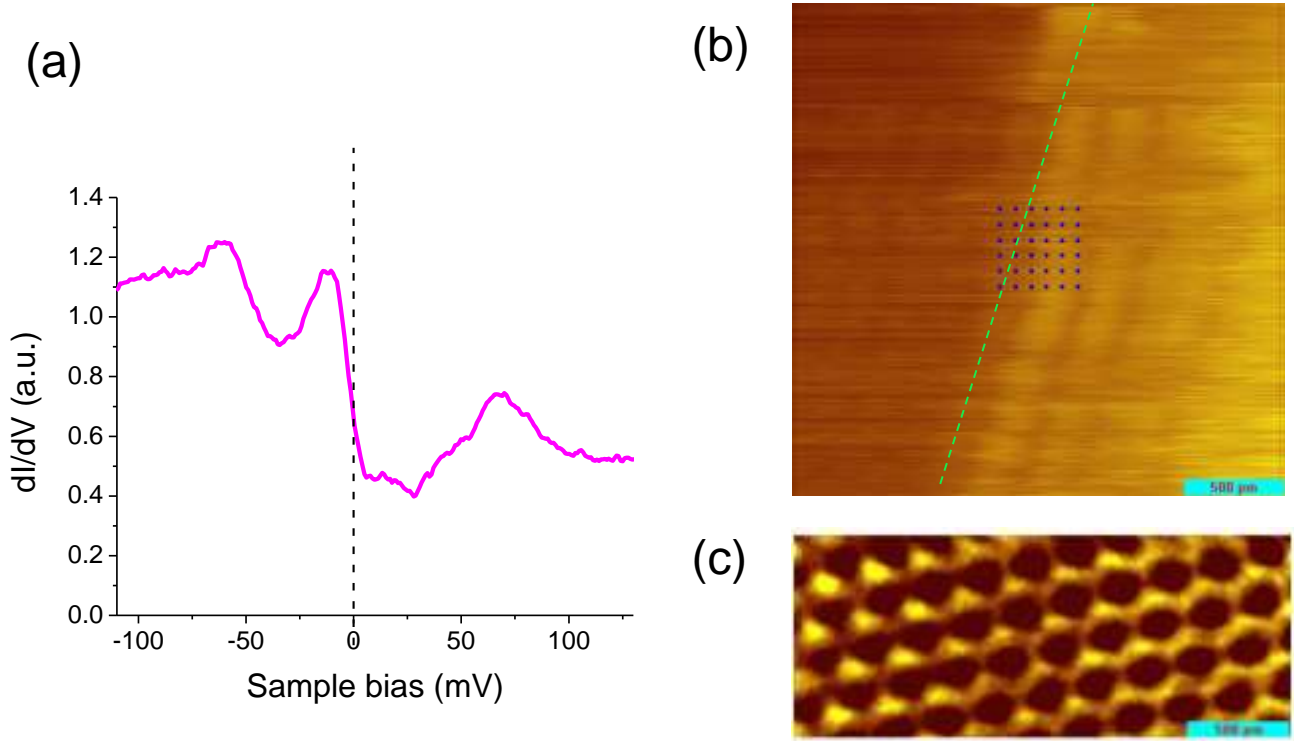

**Fig. 4. STM/STS at a zigzag edge.** (a) Spectrum is averaged over locations marked by the dots in panel (b). On the edge the spectrum is singularly different from the spectra inside the sample shown in Fig. 1c. (b) Topography at the zigzag edge (position 1 in Fig. 1b marked by red rectangle). (c) Zoom-in near position 2 in Fig. 1b (~6nm away from the edge) showing a transition from a honeycomb to triangular structure. Scale bars: 500pm in (b) and (c).